\documentclass[aps,prl,twocolumn,showpacs,titlepage,superscriptaddress]{revtex4-1}
\usepackage{amsmath}
\usepackage{graphicx}
\usepackage{hyperref}
\usepackage{float}
\hypersetup{colorlinks, linkcolor={blue},citecolor={blue}, urlcolor={blue}}
\begin{document}

\title{Experimental Observation of Resonance-Assisted Tunneling}
\author{Stefan Gehler}
\affiliation{Fachbereich Physik, Philipps-Universit\"{a}t Marburg, Renthof 5, D-35032 Marburg, Germany}
\affiliation{Department of Energy Management and Power System Operation, University of Kassel, D-34121 Kassel, Germany}
\author{Steffen L\"ock}
\affiliation{Technische Universit\"at Dresden, Institut f\"ur Theoretische Physik and Center for Dynamics, 01062 Dresden, Germany}
\affiliation{OncoRay -- National Center for Radiation Research in Oncology, Faculty of Medicine and University Hospital Carl Gustav Carus, Technische Universit\"at Dresden, Helmholtz-Zentrum Dresden-Rossendorf, Fetscherstra{\ss}e~74, PF 41,
01307 Dresden, Germany}
\author{Susumu Shinohara}
\affiliation{Max-Planck-Institut f\"ur Physik komplexer Systeme, N\"othnitzer Stra\ss{}e 38, 01187 Dresden, Germany}
\author{Arnd B\"acker}
\affiliation{Technische Universit\"at Dresden, Institut f\"ur Theoretische Physik and Center for Dynamics 01062 Dresden, Germany}
\affiliation{Max-Planck-Institut f\"ur Physik komplexer Systeme, N\"othnitzer Stra\ss{}e 38, 01187 Dresden, Germany}
\author{Roland Ketzmerick}
\affiliation{Technische Universit\"at Dresden, Institut f\"ur Theoretische Physik and Center for Dynamics 01062 Dresden, Germany}
\affiliation{Max-Planck-Institut f\"ur Physik komplexer Systeme, N\"othnitzer Stra\ss{}e 38, 01187 Dresden, Germany}
\author{Ulrich Kuhl}
\thanks{Corresponding author, email: ulrich.kuhl@unice.fr}
\affiliation{Universit\'e Nice Sophia Antipolis, CNRS, Laboratoire de Physique de la Mati\`ere Condens\'ee, UMR 7336 Parc Valrose, 06100 Nice, France.}
\affiliation{Fachbereich Physik, Philipps-Universit\"{a}t Marburg, Renthof 5, D-35032 Marburg, Germany}
\author{Hans-J\"urgen St\"ockmann}
\affiliation{Fachbereich Physik, Philipps-Universit\"{a}t Marburg, Renthof 5, D-35032 Marburg, Germany}
\date{\today}

\begin{abstract}
We present the first experimental observation of resonance-assisted tunneling,
a wave phenomenon, where regular-to-chaotic tunneling is strongly enhanced
by the presence of a classical nonlinear resonance chain.
For this we use a microwave cavity made of oxygen free copper with the shape of a desymmetrized
cosine billiard designed with a large nonlinear resonance chain in the regular region.
It is opened in a region, where only chaotic dynamics takes place, such that
the tunneling rate of a regular mode to the chaotic region increases the line width of the mode.
Resonance-assisted tunneling is demonstrated by (i) a parametric variation
and (ii) the characteristic plateau and peak structure towards the semiclassical limit.
\end{abstract}
\maketitle

Tunneling describes the possibility of a quantum particle to transmit through a barrier into a region of space,
which is inaccessible for a corresponding classical particle. It is a general wave phenomenon.
Dynamical tunneling describes the tunneling of waves between classically disjoint regions
of phase space, even without an energy barrier being present \cite{DavHel1981}.
It occurs in several variants \cite{KesSch2011}, e.g.,
from a regular region to the chaotic region
\cite{HanOttAnt1984,ShuIke1995,PodNar2003,BaeKetLoeSch2008,BaeKetLoeRobVidHoeKuhSto2008,ShuIke2012,MerLoeBaeKetShu2013},
from a regular region via the chaotic region to another regular region
\cite{LinBal1990,BBEM93,TomUll1994,DorFri1995,DemGraHeiHofRehRic2000,SteOskRai2001,HenHafBroHecHelMcKMilPhiRolRubUpc2001},
or between two chaotic regions \cite{CreWhe1999,Gut2007,DieGuhGutMisRic2014}.
It is essential for applications in
atomic and molecular physics \cite{ZakDelBuc1998,Kes2005b,WimSchEltBuc2006},
ultracold atoms \cite{HenHafBroHecHelMcKMilPhiRolRubUpc2001,SteOskRai2001,ShresthaNiLamSummyWimberger2013},
optical cavities \cite{HacNoe1997,BaeKetLoeWieHen2009,ShiHarFukHenSasNar2010,YanLeeMooLeeKimDaoLeeAn2010,SonGeRedCao2012},
quantum wells \cite{FroWilHayEavSheMiuHen2002},
and microwave resonators \cite{DemGraHeiHofRehRic2000,BaeKetLoeRobVidHoeKuhSto2008}.

\begin{figure}[H]
\includegraphics[width=.99\linewidth]{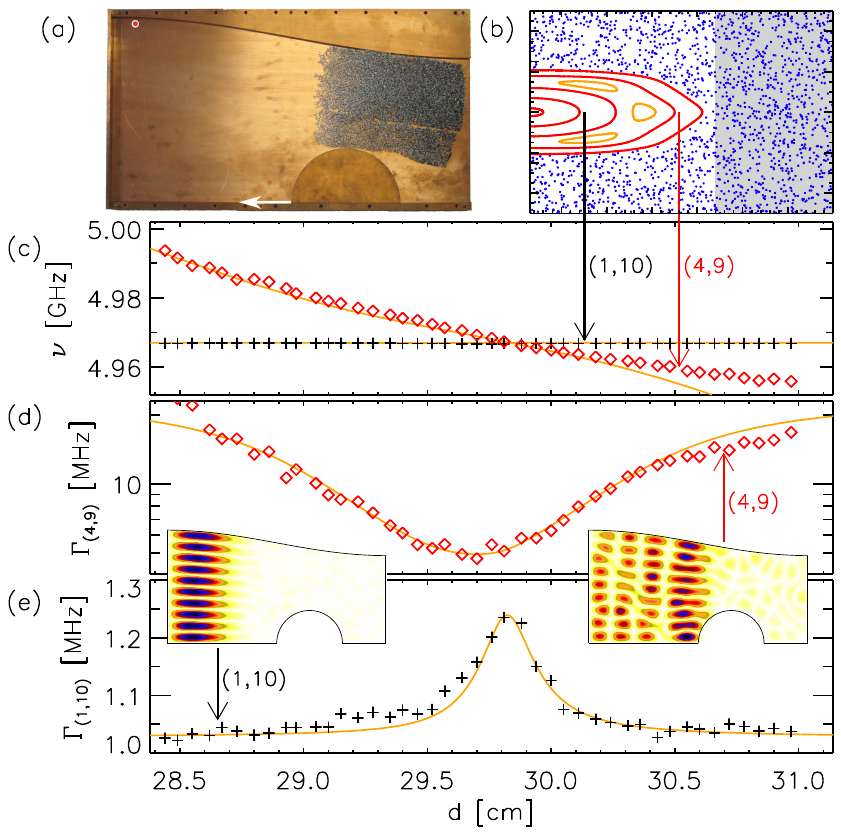}
\caption{\label{fig:n1m10}(color online).
(a) Photograph of the cavity including a movable (white arrow) half disk, absorbers (gray), and antenna position (red).
(b) Poincar\'{e} section along the upper boundary of the classical dynamics of the closed billiard
with regular tori (red), a chaotic orbit (blue), and a $3$:$1$ nonlinear resonance chain (orange). The gray region indicates the absorbing part.
(c) Frequency $\nu$ and (d), (e) width $\Gamma$
of the regular modes $(n,m)=(1,10)$ (black pluses) and $(4,9)$ (red diamonds) vs.\ half disk position $d$.
Solid orange lines correspond to the $3\times 3$ matrix model, Eq.~(\ref{eq:H3x3}).
Insets visualize numerically determined regular modes of the closed billiard.
}
\end{figure}

We consider regular-to-chaotic tunneling,
where the tunneling rate $\gamma$ describes the decay $|\psi_\text{reg}(t)|^2\propto\exp(-\gamma t)$
of a quantum state, initially located within the regular region, to the chaotic region.
Towards the semiclassical limit $\gamma$ is determined by two main effects:
For small wave numbers direct regular-to-chaotic tunneling typically leads to an exponential
decrease of $\gamma$ with increasing wave number \cite{HanOttAnt1984,BaeKetLoeSch2008,BaeKetLoe2010}.
For larger wave numbers resonance-assisted tunneling (RAT) drastically enhances the tunneling rates, causing the characteristic plateau and peak structures \cite{BroSchUll2002,EltSch2005,SheFisGuaReb2006,LoeBaeKetSch2010}.
RAT occurs due to nonlinear resonance chains inside a regular region,
see Fig.~\ref{fig:n1m10}(b) (orange lines),
which arise due to the Poincar\'e-Birkhoff theorem.
A combined prediction for direct regular-to-chaotic tunneling and RAT is given in Ref.~\cite{LoeBaeKetSch2010}.

Several experiments were performed demonstrating
direct regular-to-chaotic tunneling
\cite{DemGraHeiHofRehRic2000,HenHafBroHecHelMcKMilPhiRolRubUpc2001,SteOskRai2001,FroWilHayEavSheMiuHen2002,BaeKetLoeRobVidHoeKuhSto2008,
ShiHarFukHenSasNar2010,YanLeeMooLeeKimDaoLeeAn2010,SonGeRedCao2012,ShresthaNiLamSummyWimberger2013}.
Recently, the coupling matrix element between two modes coupled by a nonlinear resonance chain
was very nicely observed experimentally in the near integrable
regime \cite{KwaShiMooLeeYanAn2015} in microcavities \cite{YanLeeMooLeeKimDaoLeeAn2010}.
The experimental observation of the enhancement of
regular-to-chaotic tunneling rates due to RAT, however, has remained open.

\begin{figure}
\includegraphics[width=0.98\columnwidth]{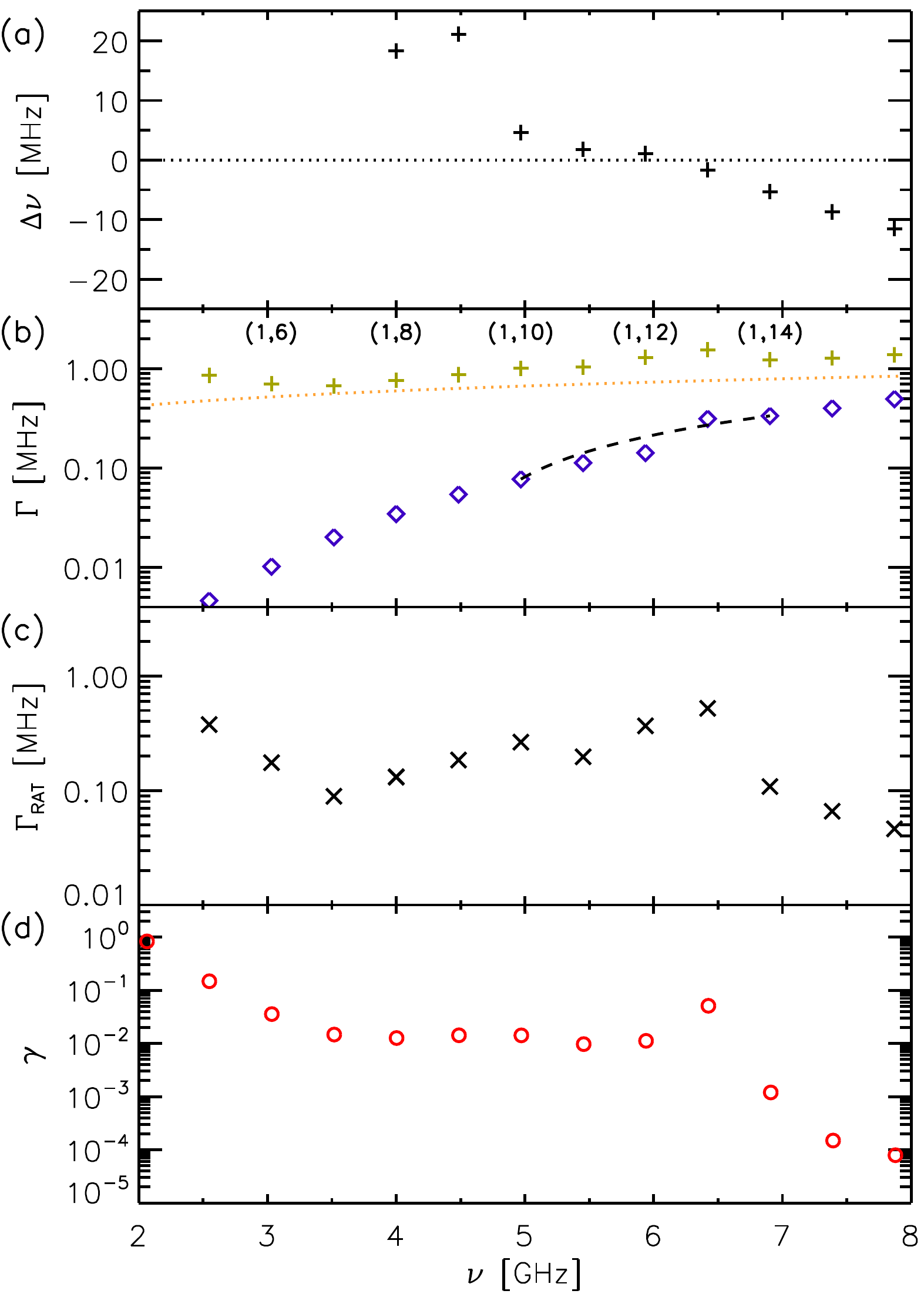}
\caption{\label{fig:series} (color online).
(a) Frequency difference $\Delta\nu=\nu_{(1,m)}-\nu_{(4,m-1)}$ of modes $(1,m)$ and $(4,m-1)$
for $m=8,\dots,16$ and fixed half disk position $d=30.0$\,cm.  The dotted line indicates the value 0.
(b) Experimental widths $\Gamma_{(1,m)}$ (green pluses) of the inner mode for $m=5,\dots,16$,
the square root dependence of $\Gamma_{\text{wall}}$ (orange dotted line),
and $\Gamma_{\text{antenna}}$ (blue diamonds) with
the dashed line showing the assumed linear dependence at the close encounter.
(c) $\Gamma_{\text{RAT}}$ (black crosses)
showing the initial exponential decay and the characteristic plateau and peak structure of RAT.
(d) Numerical dimensionless tunneling rates $\gamma_{(1,m)}$ of the closed system (red open circles).
}
\end{figure}

Here we measure the enhanced line width of regular modes
due to RAT in an open microwave cavity; see Fig.~\ref{fig:n1m10}(a).
Couplings between different regular modes are caused by a large $3$:$1$ nonlinear resonance chain; see Fig.~\ref{fig:n1m10}(b).
We show the influence of RAT in two ways:
(i) We induce RAT parametrically by a crossing of the frequencies of two regular modes
under variation of the position of a half disk inset,
see Figs.~\ref{fig:n1m10}(c)-\ref{fig:n1m10}(e); (ii) we explore the dependence of the width of regular modes
towards the semiclassical limit, i.e., for increasing frequency.
The characteristic plateau and peak structure of RAT is observed showing a good qualitative
agreement with numerically determined tunneling rates using the closed system; see Fig.~\ref{fig:series}.

For the experimental realization we use a desymmetrized cosine \cite{BaeSchSti1997,BaeManHucKet2002}
microwave resonator \cite{StoSte1990,Sri1991,GraHarLenLewRanRicSchWei1992,SoAnlOttOer1995,Sto1999},
with a large $3$:$1$ nonlinear resonance chain; see Fig.~\ref{fig:n1m10}(a).
The area enclosed by the resonance chain is approximately half of the area of the regular island, such that the characteristic plateau and peak structure of RAT
is observable on top of the experimental line width.
Figure~\ref{fig:n1m10}(b) shows a Poincar\'{e} section
with perpendicular momentum $|p_\parallel|/p \leq 1$ vs.\ arclength along the upper boundary of the cosine billiard
consisting of regular tori, a chaotic orbit, and the $3$:$1$ nonlinear resonance chain.
In the resonator (height $h=1$\,cm, length $l=60$\,cm, width left $w_l=30.96$\,cm, and right $w_r=24$\,cm)
we placed a half disk (radius $r=9$\,cm) at the bottom boundary with an adjustable distance $d$ from the left boundary, see Fig.~\ref{fig:n1m10}(a).
It destroys the hierarchical island structure and partial barriers in the chaotic region leading to a sharp separation of regular and chaotic dynamics. This reduces
strong fluctuations of the tunneling rates, which would hinder the observation of RAT.
Chaotic trajectories spread over the whole billiard also hitting the half disk, whereas the regular motion is confined to the region left of the half disk.
The semiclassical eigenfunction hypothesis \cite{Per1973,Ber1977,Vor1979} predicts regular and chaotic eigenmodes of the billiard.
Numerical examples of regular modes are shown in Figs.~\ref{fig:n1m10}(d) and ~\ref{fig:n1m10}(e) (insets).
In phase space they localize on quantizing tori
at the starting point of the arrows in Fig.~\ref{fig:n1m10}(b), as can be visualized by a Husimi representation~\cite{BaeManHucKet2002}.

We open the system by introducing a broad band foam absorber on the right-hand side [grey shaded region in Fig.~\ref{fig:n1m10}(b)]
thus connecting the chaotic part to the continuum. The surviving chaotic modes concentrate on the remaining part of the chaotic region
to the left of the absorber.
This enables the observation of regular-to-chaotic tunneling by measuring the line width of a regular mode. A tedious extraction by analyzing
many avoided crossings is not necessary \cite{BaeKetLoeRobVidHoeKuhSto2008}.

The fixed antenna is placed at the upper left part of the billiard [red dot in Fig.~\ref{fig:n1m10}(a)],
guaranteeing a good excitation of all regular modes.
The complex reflection amplitude $S_{11}(\nu)$ was measured as a function of the microwave
frequency $\nu$. In the isolated resonance regime it can be described by the Breit-Wigner formula
\begin{equation}
S_{11}(\nu)= 1 - i \sum_{k} \frac{a_{k}}{\nu-\nu_{k}-\Delta_{k}+\frac{i}{2}\Gamma_{k}},
\label{eq:BillBreitWig}
\end{equation}
where the sum runs over all modes with eigenfrequency $\nu_k$ and eigenwidth $\Gamma_{k}$.
Here $a_k=\text{Re}(\lambda_k) \psi_k(\vec{r}_1)\psi_k(\vec{r}_1)$
is the residuum of the resonance, $\psi_k(\vec{r}_1)$ the $k$th wave function
at the antenna position $\vec{r}_1$, and $\lambda_k$ the coupling coefficient of the antenna.
The real part of $\lambda_k$ describes the coupling to the
continuum and the imaginary part induces a frequency shift $\Delta_k$.
We will use the term mode for these resonances and reserve further on the term {\it resonance} to the nonlinear resonance chain.

The regular modes are labeled by two quantum numbers $(n,m)$,
where $n$ counts the number of excitations in the horizontal and
$m$ in the vertical direction.
Their width $\Gamma_{(n,m)}$ contains different contributions
\begin{equation}
\label{eq:width}
\Gamma_{(n,m)} = \Gamma_{\text{RAT}} + \Gamma_{\text{wall}} + \Gamma_{\text{antenna}},
\end{equation}
where $\Gamma_{\text{RAT}}$ is the width enhancement due to regular-to-chaotic tunneling
in the presence of the absorber.
The width $\Gamma_{\text{wall}}$ is induced by the bottom and top plates and the side walls. The width induced by the antenna, $\Gamma_{\text{antenna}} = 2 |a_{(n,m)}|$,
is given by the residuum $a_{(n,m)}$.

{\em (i) RAT induced by parameter variation.}---We first show the signature of RAT from
parametric dependencies of two close-by regular modes.
The modes $(n,m) = (1,10)$
and $(n', m') = (4,9)$ (Fig.~\ref{fig:n1m10} insets) are coupled via to the nonlinear $3$:$1$ resonance chain,
as $n' = n+3$ and $m' = m-1$.
We measure $S_{11}(\nu)$ for several half disk positions $d$, fixing all other parameters.
We fitted their frequency $\nu$ and width $\Gamma$,
giving the parametric dependencies shown in Figs.~\ref{fig:n1m10}(c)-\ref{fig:n1m10}(e).
We observe a crossing of the frequencies and an avoided crossing of the widths. An increase of the
width of the mode $(1,10)$, lying deep inside the regular region, is observed once
the mode $(4,9)$ is close by, which is a clear signature of RAT.

A quantitative description of all features displayed in Fig.~\ref{fig:n1m10} can be obtained in terms of a matrix model.
We start with a $2\times 2$ matrix description of the two regular modes with diagonal elements $E_1-i\gamma_1/2$ and $E_4(d) -i\gamma_4/2$, where $E_1$ and $E_{4}$ are the eigenenergies of the uncoupled modes, $(1,10)$ and $(4,9)$.
As the inner mode concentrates away far off the half disk, we assume $E_1$ to be independent of $d$.
Close to the crossing we assume a linear dependence on the frequency axis of $E_4(d)$ as is suggested by Fig.~\ref{fig:n1m10}(c).
The widths $\gamma_{1}$ and $\gamma_{4}$ are assumed to be independent of $d$ as they are due to antenna coupling and wall absorption.
An off-diagonal matrix element $V_{3:1}$ describes the coupling via the 3:1 nonlinear resonance chain.
The line width of mode 1, obtained by diagonalizing the $2\times 2$ matrix, increases at the point of degeneracy of the real parts of the two modes, as exhibited in Fig.~\ref{fig:n1m10}(e).
This is a manifestation of the coupling of two regular modes via a nonlinear resonance chain.

For a full analysis of RAT, however, one additionally has to consider the tunneling into the chaotic region.
If the two-mode model were the full truth, the curves in Figs.~\ref{fig:n1m10}(d) and \ref{fig:n1m10}(e) should be just mirror images of each other because of the invariance of the trace of the Hamilton matrix.
Thus the sum of the two line widths should be independent of $d$, which obviously is not the case.
Thus there must be (at least) one other mode from the chaotic region involved.
We assume a linear dependence of $E_\text{ch}(d)$  due to the change of the area to the left of the absorber causing an average drift in mode energies, described by a Taylor expansion up to linear order in $d$. The width $\gamma_\text{ch}$ is assumed to be constant.

Because of coupling with mode 4 its width dependence with $d$ must be essentially the mirror image of Fig.~\ref{fig:n1m10}(d), since the contribution of mode 1 to the width is only marginal. Hence the chaotic mode has to become broad at the crossing point, while at the same time the width of mode 4 becomes small. This is the phenomenology which is known in other contexts as resonance trapping~\cite{rot09,per00}. It is found whenever an effective Hamiltonian of the form $H=H_0-i WW^\dag$ is involved.
Here $H_0$ is the Hamiltonian of the unperturbed system, and $W$ is a $N\times M$ coupling matrix, where $N$ is the number of modes taken into account and $M$ is the number of open channels
(here $N=3$ and $M\approx 6$ is obtained by dividing the length of the absorbing edge by $\lambda/2$).
This suggests an ansatz of the Hamiltonian in terms of a $3\times 3$ matrix,
\begin{equation}\label{eq:H3x3}
H=\left(\begin{matrix}
E_{1}-i \dfrac{\gamma_{1}}{2} & V_{3:1} & 0 \\
V_{3:1} & E_{4}(d) - i \dfrac{\gamma_{4}}{2} & -i V_{\text{dir},4}\\
0 & -i V_{\text{dir},4} & E_\text{ch}(d) -i \dfrac{\gamma_\text{ch}}{2}
\end{matrix}\right).
\end{equation}
This ansatz assumes ($\alpha$) that the direct tunneling coupling of mode 1 to the chaotic region is negligible
for the frequency range of Fig.~\ref{fig:n1m10}
and ($\beta$) that the direct tunneling coupling of mode 4 to the chaotic region dominantly appears
in the coupling of mode 4 to the open channels encoded in the second row of $W$
(and much less in the coupling of mode 4 to the chaotic mode in $H_0$).
As the real parts of modes 1 and 4 cross, with no indication of an avoided crossing [Fig.~\ref{fig:n1m10}(c)], the off-diagonal elements of $H$ have to be either real or imaginary, but not complex.
After diagonalization of the $3\times3$ Hamiltonian
we have to compare the complex quantum mechanical eigenenergies $\tilde E = E- i {\gamma}/{2}$
with the experimentally measured complex electromagnetic eigenvalues $\tilde\nu = \nu -i\Gamma/2$
by their complex wave number $\tilde E /(\hbar^2/2M) = (2\pi\tilde\nu/c)^2$,
yielding
\begin{equation}\label{eq:E2qm}
E=\nu^2 \cdot (2\pi/c)^2;\quad \gamma=2\nu \Gamma \cdot (2\pi/c)^2,
\end{equation}
where we use units $\hbar/2M =1$ and $\Gamma \ll \nu$.
Fitting the experimental data gives the solid orange lines in Figs.~\ref{fig:n1m10}(c)-\ref{fig:n1m10}(e)
and yields the parameters of the model~\footnote{The fit yields:
$\nu_{(1,10)}=4.967$\,GHz,
$\Gamma_{(1,10)}=1.03$\,MHz,
$\nu_{4}=\nu_{(1,10)}-0.023(d-29.77$\,cm$)$\,GHz/cm,
$\Gamma_{4}=23.5$\,MHz,
$\nu_\text{ch}= \nu_{(1,10)}+0.02(d-29.55$\,cm$)$\,GHz/cm,
$\Gamma_\text{ch}=70$\,MHz,
$V_{3:1}=2.3\text{m}^{-2}$,
and $V_{\text{dir},4}=75.6\,\text{m}^{-2}$.
The very broad chaotic mode is still not too far in the complex plane
due to the imperfect absorber. Assuming an absorber reflection of a few percent,
which is a reasonable value, and using a quasi-one-dimensional open rectangular billiard
the mode width can be estimated to be about $125$\,MHz, which is of the order of the fitted $\Gamma_\text{ch}$ above.
}.
In particular, this gives for the matrix element $V_{3:1}=2.3\,\text{m}^{-2}$,
which is of the same order as the prediction for the closed system, $V_{3:1,\text{cl}}=0.51\,\text{m}^{-2}$
\footnote{The matrix element $V_{3:1}$ is calculated by $V_{3:1}=E(S_+-S_-)\arccos(\text{Tr}\,M/2)/(32\cdot 6^2 \cdot l^2)$, similar as that performed in
Ref.~\cite{EltSch2005} for kicked systems.
Here $E$ is the considered energy, $S_+$ and $S_-$ are the
areas enclosed by the outer and inner separatrix of the $3$:$1$
resonance in a Poincar\'e section with unit energy,
and $\text{Tr}\,M$ is the trace of the linearized mapping of the fixed point in the center of the resonance.
With $E=10837\,\text{m}^{-2}$, $S_+=0.148\,\text{m}^{2}$, $S_-=0.108\,\text{m}^{2}$, and $\text{Tr}\,M=1.77$ one finds $V_{3:1,\text{cl}}=0.51\,\text{m}^{-2}$.
}.

We would like to emphasize that the third chaotic mode is an effective description of the coupling of the system to the environment, more precisely speaking to the absorber. For the example of Fig.~\ref{fig:n1m10} we chose a situation where the effective broad chaotic mode crosses almost at the same $d$ value where the two regular modes cross. For other quantum numbers $m$ the situation was comparable but the crossing of the chaotic state was further apart. We find a similar value for $V_{3:1}$ using a reduced $2\times2$ model, as this value is defined mainly by the width increase of $\Gamma_{(1,10)}$ close to $d=29.8$\,cm. We considered either a constant width $\Gamma_{4}$ using only the range 29.5\,cm $\le d\le 30.0$\,cm, or an adjusted cosine or Gaussian width dependence $\Gamma_{4}(d)$ for the whole range of $d$. This shows that the signature of RAT in Fig.~\ref{fig:n1m10}(e) can be quantitatively understood.

{\em (ii) RAT plateau and peak structure.}---A key feature of RAT is the plateau and peak structure on top of the exponential
decay of the tunneling rates towards the semiclassical limit.
We observe this signature of RAT with fixed half disk position $d=30.0$\,cm.
The plateau should appear when in addition to the inner mode $(1,m)$
the outer mode $(4,m-1)$ starts to exist ($m \geq 8$).
A peak is expected at the crossing of these modes
for increasing $m$, as seen in Fig.~\ref{fig:series}(a) for
the frequency difference $\Delta\nu=\nu_{(1,m)}-\nu_{(4,m-1)}$ around 6\,GHz.
There the corresponding quantizing tori in phase space are symmetric
with respect to the nonlinear resonance chain.
From the measured widths $\Gamma_{(1,m)}$ [green pluses in Fig.~\ref{fig:series}(b)]
for $m=5,\dots,16$
we subtract, according to Eq.~(\ref{eq:width}),
the width contributions $\Gamma_{\text{wall}}$ and $\Gamma_{\text{antenna}}$
(discussed below)
and the resulting contribution $\Gamma_{\text{RAT}}$ is shown as
black crosses in Fig.~\ref{fig:series}(c).
For low frequencies up to 3.5\,GHz an exponential dependence is obtained,
as expected from direct regular-to-chaotic tunneling.
RAT becomes visible from 4\,GHz on as a broad plateau ($m \geq 8$) ending in a peak structure around 6.5\,GHz.

For numerical comparison, tunneling rates were determined using the closed system
in a similar way as described in Ref.~\cite{BaeKetLoeRobVidHoeKuhSto2008}.
Extending the chaotic region by varying the height of a rectangle that was attached
to the bottom of the cosine billiard at $x>30\,\text{cm}$, tunneling rates $\gamma_{(1,m)}$ are determined
by evaluating avoided crossings between the regular mode $(1,m)$ and chaotic modes [Fig.~\ref{fig:series}(d)].
Qualitatively these rates are in good agreement with the experimental results $\Gamma_{\text{RAT}}$ in Fig.~\ref{fig:series}(c).
A quantitative comparison is difficult, as $\Gamma_{\text{RAT}}$
depends also on the wall and antenna contributions $\gamma_4$ and $\gamma_\text{ch}$
which differ for each $m$.
However, the experimental and numerical frequency ranges for the initial exponential decay, the plateau, and the peak
are identical.

Let us finally discuss the wall and antenna contributions
$\Gamma_{\text{wall}}$ and $\Gamma_{\text{antenna}}$ in Fig.~\ref{fig:series}(b).
The wall absorption $\Gamma_{\text{wall}}$ has a square root frequency dependence due to nonperfect conductance of the cavity walls \cite{Jac1962}.
It was extracted from measurements of the closed cavity, where the absorber was removed and the right part of the billiard
was closed. We find $\Gamma_{\text{wall}} = 0.30$\,MHz$\sqrt{\nu/\text{GHz}}$;
see the orange dotted line in Fig.~\ref{fig:series}(b).
The width $\Gamma_{\text{antenna}} = 2 |a_{(1,m)}|$ induced by the antenna
is shown as blue diamonds in Fig.~\ref{fig:series}(b) and is
extracted by fitting a Lorentzian to the measured resonance.
Around 6\,GHz, where the frequency difference $\Delta\nu$ in Fig.~\ref{fig:series}(a) is small,
modes 1 and 4 are coupled, such that the extracted $\Gamma_{\text{antenna}}$ is affected.
In this regime we therefore interpolate it by a linear increase; see the
dashed line in Fig.~\ref{fig:series}(b).
The overall monotonic increase of $\Gamma_{\text{antenna}}$ is due to two effects:
($\alpha$) the antenna length is small compared to the wavelength leading to small and
increasing antenna coupling $\text{Re}(\lambda)$, and
($\beta$) the distance of the antenna to the wall is smaller than a quarter of the wavelength
such that the wave function increases monotonically at the antenna position.

In this Letter, we experimentally observe RAT in a generic mixed phase space
using an opened microwave billiard.
We demonstrate RAT by the width increase of the exemplary mode $(1,10)$ when it is crossed by mode $(4,9)$.
With a $3\times 3$ matrix model the
coupling matrix element $V_{3:1}$ is extracted, which is of the same order as the theoretical value.
Second, measuring the widths of the modes $(1,m)$ for increasing $m$ reveals the
regimes of direct regular-to-chaotic tunneling with an exponential decay
and RAT with its characteristic plateau and peak structure.
This experimental work motivates future theoretical studies for a better quantitative
description of RAT, in particular using semiclassical methods~\cite{ShuIke1995,MerLoeBaeKetShu2013}.

\begin{acknowledgments}
We thank N.~Mertig for valuable discussions and
acknowledge support by the Deutsche Forschungsgemeinschaft
within the Forschergruppe 760 ``Scattering Systems with Complex Dynamics.''
\end{acknowledgments}

\end{document}